# Driving Thermoelectric Optimization in AgSbTe$_2$ via Design of Experiments and Machine Learning

*Jan-Hendrik Pöhls\*, Chun-Wan Timothy Lo, Marissa MacIver, Yu-Chih Tseng, Yurij Mozharivskyj\**

*Jan-Hendrik Pöhls, Chun-Wan Timothy Lo, Marissa MacIver, Yurij Mozharivskyj*
Department of Chemistry and Chemical Biology, McMaster University, 1280 Main Street West, Hamilton, Ontario L8S 4M1, Canada

*Jan-Hendrik Pöhls*
Department of Chemistry, University of New Brunswick, 30 Dineen Drive, Fredericton, New Brunswick E3B 9W4, Canada

*Yu-Chih Tseng*
CanmetMATERIALS, Natural Resources Canada, Hamilton, Ontario L8P 0A5, Canada

\*Corresponding authors: Email: Jan.Pohls@unb.ca
Email: mozhar@mcmaster.ca

**Abstract**

Systemic optimization of thermoelectric materials is arduous due to their conflicting electrical and thermal properties. A strategy based on Design of Experiments and machine learning is developed to optimize the thermoelectric efficiency of AgSb$_{1+x}$Te$_{2+y}$, an established thermoelectric. From eight experiments, high thermoelectric performance in AgSb$_{1.021}$Te$_{2.04}$ is revealed with a peak and average thermoelectric figure of merit of 1.61±0.24 at 600 K and 1.18±0.18 (300 – 623 K), respectively, which is >30% higher than the best literature values for AgSb$_{1+x}$Te$_{2+y}$. Ag-deficiency and suppression of secondary phases in AgSb$_{1.021}$Te$_{2.04}$ improves the electrical properties and reduces the thermal conductivity (~0.4 W m$^{-1}$ K$^{-1}$). Our strategy is implemented into an open-source graphical user interface, and it can be used to optimize the methodologies, properties, and processes across different scientific fields.

# 1. Introduction

Most of the energy produced worldwide is dissipated in the form of thermal energy. Thermoelectric materials have the ability to recover such waste heat and convert it into electrical energy. Therefore, wide-scale applications of thermoelectric materials are vital for our transition to clean energy. Despite this promise, affordable thermoelectric applications are currently limited by their low efficiencies.

The thermoelectric efficiency, $\eta = \frac{\Delta T}{T_h} \cdot \frac{\sqrt{1+\overline{zT}}-1}{\sqrt{1+\overline{zT}}+{T_c}/{T_h}}$, is related to the thermoelectric figure of merit, $zT = \frac{S^2 T}{\rho \cdot \kappa}$, averaged over operational temperatures (with $\overline{zT}$ as the average $zT$, $\Delta T = T_h - T_c$, $T_h$, and $T_c$ as the temperature difference, hot and cold side absolute temperatures, respectively, $S$ as the Seebeck coefficient, $\rho$ as the electrical resistivity, and $\kappa$ as the total thermal conductivity).[1] The total thermal conductivity can be further split into electronic, $\kappa_{el}$, and phononic, $\kappa_{ph}$, contributions. All electrical and thermal properties are intertwined, and adjusting a single parameter often leads to an unfavorable offset in the other properties, significantly reducing the thermoelectric performance. While $|S|$ and $\rho$ decrease with increasing charge carrier concentration, $\kappa_{el}$ increases.[1] Similarly, a higher band effective mass improves $|S|$, but also increases $\rho$ and, hence, it may reduce the power factor, $PF = S^2/\rho$.[2]

Over the last few decades, new methods and strategies have emerged to drive the optimization of thermoelectric materials by enhancing electrical properties[3,4,5] or reducing $\kappa_{ph}$.[6,7] For instance, multiscale hierarchical structuring can reduce $\kappa_{ph}$ by efficiently scattering phonons at various length scales, while appropriate doping can improve the power factor by adjusting the carrier concentration and optimizing the band structure. These concepts have led to the development of new materials with $zT$s values more than doubled those 20 years previously.[8] However, optimization of thermoelectric materials is challenging as subtle changes in the chemical compositions can drastically affect the thermoelectric performance. For example, the carrier concentration of In-doped $PbTe_{0.996}I_{0.004}$ is very sensitive to the indium content, resulting in a high thermoelectric performance for lightly doped $In_{0.0035}Pb_{0.9965}Te_{0.996}I_{0.004}$.[9]

In recent years, computational methods have been rigorously applied to optimize thermoelectric materials. Recent density functional theory (DFT) efforts have accurately predicted the electrical[10] and thermal[11] properties of thermoelectric materials.[12] However, the thermoelectric properties of semi-metals and narrow-gap semiconductors in which electrons and holes both contribute to the charge transport cannot be accurately estimated. Furthermore, the prediction of doped samples is limited due to high computational cost. Subtle changes in the composition alter the band structures and relaxation times. Therefore, finding the optimum composition is elusive, most notably when the compositions contain multiple elements to adjust.

Experimentally, optimizing multiple variables (*i.e.*, Sb and Te concentrations in the $AgSb_{1+x}Te_{2+y}$ system) is also arduous. In the traditional One-Variable-At-a-Time (OVAT) approach, it is often assumed that all variables are independent and therefore, each variable can be separately screened and in an arbitrary order.[13] Previously optimized variable(s) act as a starting point for orthogonal screening of the other variable(s). Although this approach can improve the desired properties, the optimum value often will not be found because most variables are correlated, such as ρ, *S*, and κ in thermoelectric materials. Furthermore, the conventional approach requires many data points to optimize one variable at a time. The measurement of thermoelectric properties is time-consuming, limiting the number of experiments and hence, the OVAT approach is not ideal for optimization of thermoelectric efficiency.

Design of Experiments (DoE) approach can speed up the optimization process and is often used in pharmacy, biology, and chemical engineering.[13,14,15,16] In one study, the efficiency of organic photovoltaics has been successfully optimized using DoE and machine learning (ML).[15] However, a large number of experiments [~29] were still conducted to optimize the efficiency. To our knowledge, DoE combined with ML has never been applied to thermoelectric materials or other solid-state materials. In this study, we developed software combining Design of Experiments and machine learning to accelerate the optimization of the thermoelectric efficiency (**Figure 1(b)**). Our approach was successfully tested on the $AgSb_{1+x}Te_{2+y}$ system – a narrow band semiconductor ($E_g$ = 0.01 – 0.38 meV) [17,18] with high thermoelectric performance.[7,19,20,21,22] In just three optimization cycles and in total eight samples, we increased the maximum ($zT_{peak}$ = 1.61±0.24) and average ($\overline{zT}$ = 1.18±0.18) by more than 30% compared to the best published data, $zT_{peak}$ = 1.22 and $\overline{zT}$ = 0.9 (**Figure 1(c) and (d)**).[19,20]

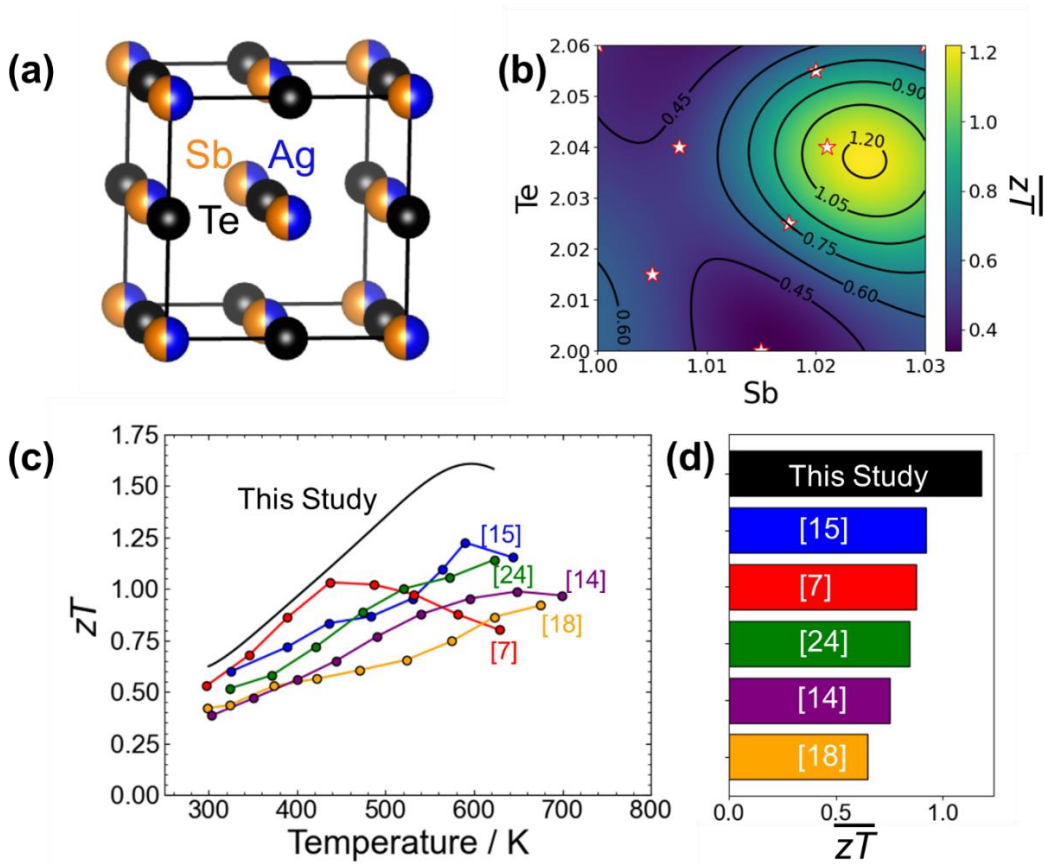

**Figure 1. Thermoelectric optimization of AgSb$_{1+x}$Te$_{2+y}$ via Design of Experiments (DoE) and machine learning (ML).** (a) Disordered rock-salt structure of AgSbTe$_2$. (b) Machine learning-driven thermoelectric optimization using a support vector regression model. Stars indicate the chemical compositions of the synthesized samples predicted by the DoE approach. Comparison of (c) the thermoelectric figure of merit, *zT*, and (d) average thermoelectric figure of merit, $\overline{zT}$, of the DoE/ML optimized AgSb$_{1+x}$Te$_{2+y}$ compounds with the previously reported compounds.[7,18, 19, 20, 21, 22,28] Note, $\overline{zT}$ is the average *zT* from 300 K to 623 K.

## 2. Results and Discussion

### 2. 1. Background on AgSb$_{1+x}$Te$_{2+y}$

AgSbTe$_2$ crystallizes in a disordered rock salt-type crystal structure in which Ag$^+$ and Sb$^{3+}$ ions randomly occupy the cation site (**Figure 1(a)**). Pristine AgSbTe$_2$ possesses extremely low room-temperature κ (0.6 – 0.7 W m$^{-1}$ K$^{-1}$),[23,24,25] which is attributed to strong anharmonicity,[25] low-frequency optical modes,[26] and nanoscale domains with short-range cations ordering.[27] The low κ, and the high *PF* of pristine AgSbTe$_2$ due to a high band degeneracy results in an impressive *zT*. However, thermoelectric efficiency of pristine AgSbTe$_2$ is limited by the cation disorder and impurity phases (*e.g.*, Ag$_2$Te).[7] Extrinsic dopants (*e.g.*, In, Ce, Zn) can reduce the formation of impurity phases in AgSbTe$_2$, improving $zT$.[18,19,20,21,22] In a recent study, Cd-doping was reported to supress both the cation disorder and impurity precipitation, resulting in a simultaneous increase of the electrical properties and reduction of the thermal conductivity, leading to an extraordinary thermoelectric performance in the medium temperature range ($\overline{zT} \sim 1.9$ from 300 to 573 K).[7] Impurity phases in the AgSb$_{1+x}$Te$_{2+y}$ system also can be decreased by rapid cooling.[28] Ag$_2$Te precipitations dissolve in the AgSbTe$_2$ matrix above ~630 K and faster cooling reduces the formation of Ag$_2$Te resulting in a higher $\overline{zT}$. However, the quenched sample had lower thermoelectric performance than the slowly cooled one indicating that other parameters have to be considered.[7]

Our computations indicate that a subtle change in composition can drastically affect the electrical properties (**Figure S1**). Those properties were calculated using the semi-classical Boltzmann transport and constant relaxation time approach (**Figure S1(b)**). While the *PF* nearly vanishes at the Fermi energy, it becomes substantial for the *p*-doped AgSbTe$_2$ (at ~0.9 eV below the Fermi energy). To lower the Fermi level deep into the valence band, Ag vacancies can be easily formed due to its weak binding energy.[19] Additionally, Ag-poor compounds have less of the α-Ag$_2$Te secondary phase.[21] However, it was also predicted that Ag vacancies reduce the *PF*,[29] and a precise control of the composition is necessary to optimize thermoelectric performance in AgSb$_{1+x}$Te$_{2+y}$.

## 2.2. The DoE + ML Optimization Process

To speed up the optimization process and made it more rigorous, we have developed a graphical user interface (GUI) "**M**ulti-variable **O**ptimization software driven by **D**esign of **E**xperiments and **M**achine learning (**MODEM**)" and made freely available for a wide scientific and research community. It has the advantage that no programming skills are required, an automated flow can be easily followed, and the results can be quickly revealed.[30]

In the first step of the automated flow, the AgSb$_{1+x}$Te$_{2+y}$ compositions were predicted using the Latin square approach in the DoE framework (**Figure 2 (a)**). In the Latin Square method, each parameter value can only appear once for the entire set of experiments. It is important to restrict the number of experiments and the screening design space by choosing the minima, maxima, and intervals of each variable. To cover the entire screening design space, the distances between the individual experiments were maximized using a Monte-Carlo approach. The parameter values are randomly assigned for each experiment. If the Latin square condition is fulfilled, the standard deviation of the distances between each experiment divided by the minimum distance was compared to the previous value and if smaller, replaced. This was repeated at least 10,000,000 times to optimize the coverage of the entire space.

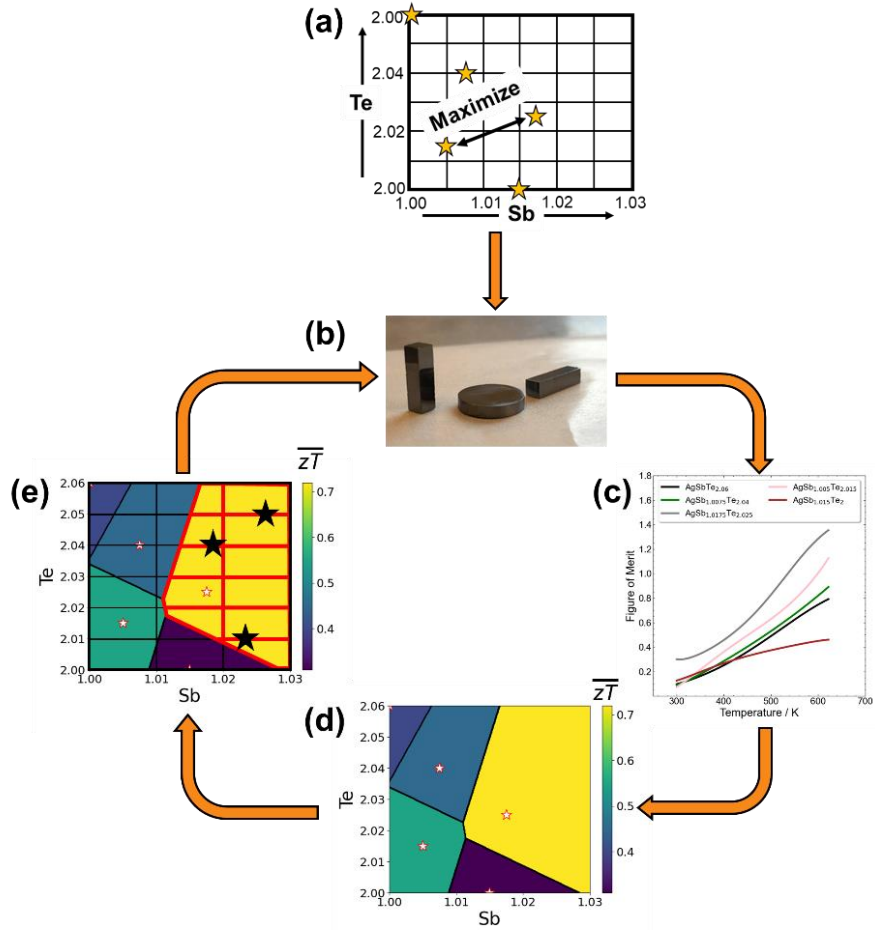

**Figure 2. Design of Experiments (DoE) and machine learning (ML) cycle to optimize thermoelectric performance.** (a) Multiple optimal compositions were predicted using the Latin square approach in the DoE framework. (b) The samples were synthesized, and (c) the thermoelectric properties were measured. (d) A support vector classification algorithm was applied to reveal the optimum area (yellow) which is (e) a constraint for the DoE algorithm. The cycle (b-e) is repeated until the optimum composition is found. White and black stars are previous and new predicted chemical compositions, respectively, to synthesize and analyze.

After the compositions had been predicted, the samples were synthesized using a solid-state method and consolidated to dense pellets (≥97% of the theoretical density) using spark plasma sintering (**Figure 2(b)**). (Details about the synthesis and characterization can be found in the Supplementary Information [SI].) The electrical and thermal properties were measured as a function of temperature and used to determine $zT$ (**Figure 2(c)**). The composition and the corresponding $\overline{zT}$ (averaged from 300 K to 623 K) were fed into a (ML) support vector classification algorithm to predict the area with the highest thermoelectric performance; see yellow area in **Figure 2(d)**. This area was fed into the DoE Latin Square approach as a new constraint for the Monte-Carlo algorithm (**Figure 2(e)**). To accelerate the DoE approach, the limits of each variable was set to the limits of the optimized area. Furthermore, the interval was accordingly reduced to randomly sample the new space. The cycle was repeated until the maximum thermoelectric performance was reached.

## 2.3. Thermoelectric Performance of AgSb$_{1+x}$Te$_{2+y}$

The thermoelectric efficiency of AgSb$_{1+x}$Te$_{2+y}$ was optimized in three cycles (**Figure 3**). In the first optimization cycle, five different AgSb$_{1+x}$Te$_{2+y}$ ($0 \leq x \leq 0.02$; $0 \leq y \leq 0.06$) compositions were proposed from DoE. While most of the synthesized samples had some of the Ag$_2$Te secondary phase, AgSb$_{1+x}$Te$_{2.06}$ samples contained Sb$_2$Te and Ag$_5$Te$_3$ impurity phases (**Figure S2**). The temperature-dependent $zT$ reveals a large spread in the thermoelectric properties (**Figure 3(a)**). Low thermoelectric performance was determined for pure Sb-rich (AgSb$_{1.015}$Te$_2$) and Te-rich (AgSbTe$_{2.06}$) compounds with $zT_{peak}$=0.46±0.07 and 0.80±0.12, respectively. However, the $zT_{peak}$=1.36±0.20 achieved for AgSb$_{1.0175}$Te$_{2.025}$ was 10% better than the highest value reported previously.[19,20] The figure of merit for all compositions increased with temperature, predicting that $zT$ will be significantly higher with rising temperature. Support vector regression indicated that a simultaneous excess of Sb and Te would improve the thermoelectric efficiency, *i.e.*, samples in which the ratio of Sb to Te excess is close to 2:3 (as in AgSb$_{1.0175}$Te$_{2.025}$) would have better performance (**Figure 3(b)**). As for $zT_{peak}$, $\overline{zT}$ is more than doubled by slightly changing the compositions (**Table 1**). The area of the optimum efficiency was predicted via support vector classification (**Figure 3(c)**). It is important to note that the area is large and indefinitely increases with Sb and Te contents. Therefore, the screening design space was increased to $1.00 \leq$ Sb $\leq 1.03$ to include highly Ag-deficient compositions.

Table 1: Thermoelectric properties of AgSb$_{1+x}$Te$_{2+y}$ averaged from 300 to 625 K.

| Cycle | Composition | $\bar{\rho}$ / mΩ·cm | $\bar{S}$ / μV·cm$^{-1}$ | $\bar{\kappa}$ / W·m$^{-1}$·K$^{-1}$ | $\overline{zT}$ |
|---|---|---|---|---|---|
| 1 | AgSbTe$_{2.06}$ | 1.61±0.08 | 103±5 | 0.77±0.04 | 0.41±0.06 |
|   | AgSb$_{1.005}$Te$_{2.015}$ | 2.50±0.13 | 139±7 | 0.67±0.03 | 0.56±0.08 |
|   | AgSb$_{1.0075}$Te$_{2.04}$ | 1.90±0.10 | 110±6 | 0.68±0.03 | 0.46±0.07 |
|   | AgSb$_{1.015}$Te$_2$ | 21.5±1.1 | 277±14 | 0.54±0.03 | 0.31±0.05 |
|   | AgSb$_{1.0175}$Te$_{2.025}$ | 4.49±0.22 | 190±10 | 0.53±0.03 | 0.73±0.11 |
| 2 | AgSb$_{1.02}$Te$_{2.055}$ | 1.33±0.07 | 118±6 | 0.70±0.04 | 0.73±0.11 |
|   | AgSb$_{1.03}$Te$_{2.06}$ | 2.16±0.11 | 140±7 | 0.80±0.04 | 0.59±0.09 |
| 3 | AgSb$_{1.021}$Te$_{2.04}$ | 6.42±0.32 | 256±13 | 0.41±0.02 | 1.18±0.18 |

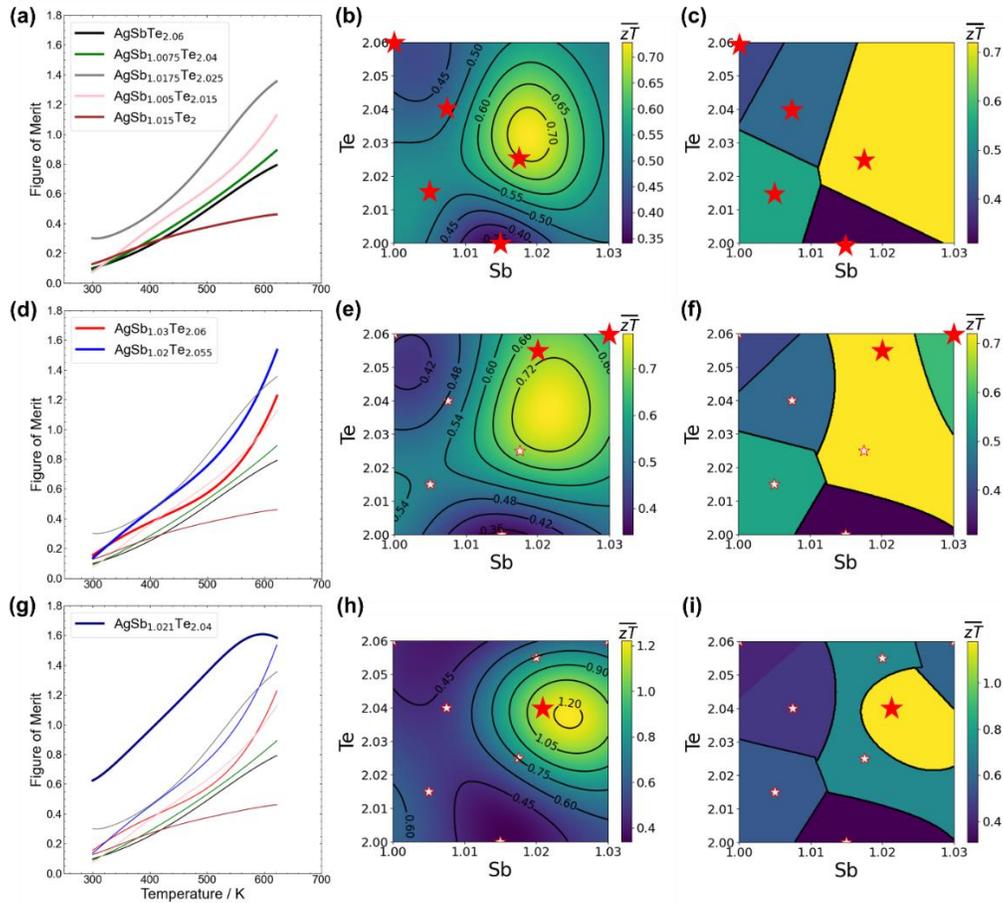

**Figure 3. Thermoelectric optimization cycles of AgSb$_{1+x}$Te$_{2+y}$.** Temperature dependence of the figure of merit for the (a) first, (d) second, and (g) third optimization cycles. (b, e, h) Regression and (c, f, i) classification of the average figure of merit using a support vector machine algorithm. Optimum (yellow) area from classification was applied as a constraint to the Design of Experiments algorithm, circulating into the optimum thermoelectric efficiency of AgSb$_{1+x}$Te$_{2+y}$. Note, $zT$ has an uncertainty of 15%. White and red stars are previous and new data points, respectively, for the support vector regression algorithm.

Both compositions in the second cycle showed high thermoelectric performance, and $zT_{peak}$ increased to 1.54±0.23 for AgSb$_{1.02}$Te$_{2.055}$ (**Figure 3(d)**). However, the thermoelectric efficiency did not increase compared to the first cycle (**Table 1**). Although higher $\overline{zT}$ was predicted for samples with extra Sb and Te, high Ag deficiency led to a decrease in thermoelectric efficiency instead (**Figure 3(e)**).

The optimum area decreased which limited the potential compositions for the third cycle (**Figure 3(f)**). Therefore, only one composition (AgSb$_{1.021}$Te$_{2.04}$) was prepared. There was no evidence of Ag$_2$Te or other secondary phases in the sample. However, the compound was composed of two AgSb$_{1+x}$Te$_{2+y}$ phases with similar lattice constants (**Figure S3** and **Table S1**). AgSb$_{1.021}$Te$_{2.04}$ exhibited higher $zT$ over the entire range compared to other compositions (**Figure 3(g)**) and previously reported AgSb$_{1+x}$Te$_{2+y}$ compounds (**Figure 1(c)**). The peak and average $zT$s values of 1.61±0.24 (at ~600 K) and 1.18±0.18, respectively, are more than 30% higher than previous reports for the pristine samples (**Figure 1**), and the present $\overline{zT}$ is higher than those of the Pb ($\overline{zT}$=0.58),[31] Sn ($\overline{zT}$=0.63),[32] In ($\overline{zT}$=1.02),[18] Se ($\overline{zT}$=1.03),[20] and Na ($\overline{zT}$=1.10),[19] doped samples and comparable to that of Zn ($\overline{zT}$=1.30)[21] doped sample. Although the thermoelectric performance of the Cd-doped sample is higher ($\overline{zT}$=1.90),[7] the present DoE/ML approach can be easily adapted to improve the thermoelectric performance of doped samples by adding a third dimension to the DoE algorithm. Furthermore, slight adjustments to the compositions can increase $\overline{zT}$ further (**Figure 3(h)**) by using the reduced area of optimum efficiency (**Figure 3(i)**).

## 2.4. Transport properties of AgSb$_{1+x}$Te$_{2+y}$

In addition to the thermoelectric performance, the temperature-dependent electrical and thermal properties are strongly dependent on composition (**Figure 4**). Most of the resistivities marginally increase with temperature suggesting heavily doped semiconducting behavior (**Figure 4(a)**). With increasing resistivity, the trend changes to nearly temperature-independent; the resistivity shows a maximum suggesting the onset of bipolar conduction. The support vector regression indicates that low average electrical resistivity is observed for the anion-rich compounds whereas the highest average resistivity is found for the Sb-rich compounds (**Figure 4(b)**). The average resistivity is 1.33 - 2.16 mΩ·cm for the Te-rich samples, which is an order of magnitude lower than the average value for AgSb$_{1.015}$Te$_2$ (**Table 1**). Most samples included Ag$_2$Te as a secondary phase. This is disadvantageous as its *n*-type behavior increases the resistivity.[21] Other secondary phases ($Sb_2Te$ and $Ag_5Te_3$) in the Te-rich compounds lead to a decrease in the resistivity.

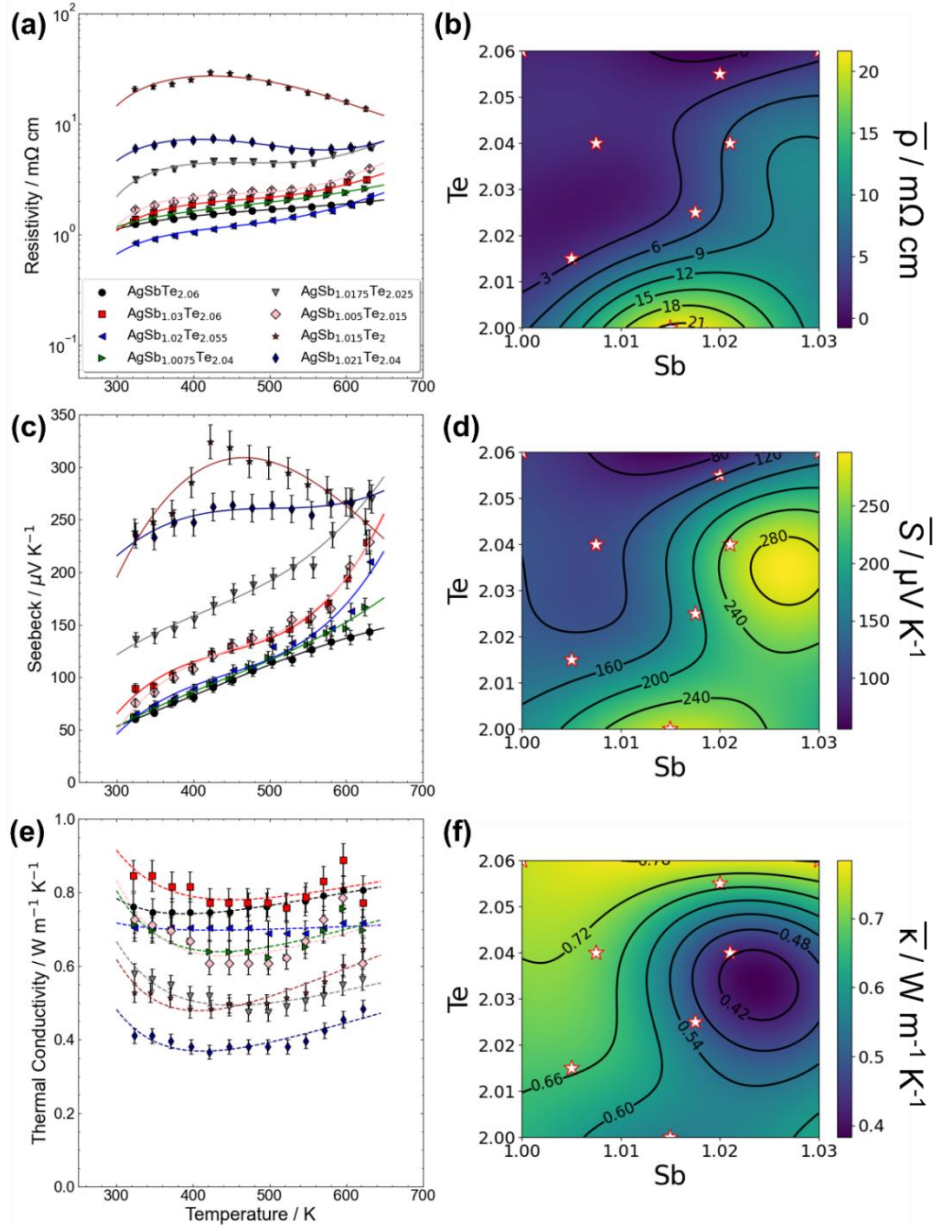

**Figure 4. Trends of electrical and thermal properties of $AgSb_{1+x}Te_{2+y}$.** (a) Temperature-dependent electrical resistivity indicates semi-metallic behavior. Lines are for visual guidance. (b) Support vector regression of the average resistivity shows that the anion-rich samples exhibit lower resistivity than the cation-rich compounds. (c) Seebeck coefficients reveal *p*-type behavior with a Goldsmid-Sharp band gap of 0.27 eV. (d) A higher average Seebeck coefficient results for the Sb-rich compounds. (e) The ultralow thermal conductivity of all compounds is nearly temperature independent. (f) Comparing the average thermal conductivity with the average resistivity indicates that the electronic and phononic contributions both are dependent on the chemical composition.

The temperature-dependent Seebeck coefficients have a trend similar to the resistivity and are positive indicating holes as the dominant charge carriers (**Figure 4(c)**). While the Seebeck coefficient for most compounds increased with temperature, the Seebeck coefficient of AgSb$_{1.015}$Sb$_2$ peaked at ~425 K, suggesting that both holes and electrons contribute to the Seebeck coefficient. The band gap estimated from the Goldsmid-Sharp formula ($E_g = 2e\cdot|S_{max}|\cdot T_{max}$) is 0.27 eV which is similar to the experimental band gap from with diffuse reflectance.[18] The Seebeck coefficient ranged from 50 – 230 µV K$^{-1}$, which is slightly lower than the reported values (160 – 280 µV K$^{-1}$).[7,18] In contrast to the electrical resistivity, the average Seebeck coefficient has two optimum regions: Sb-rich and Sb-rich/partially Te-rich (**Figure 4(d)**). An increase in Sb lowers the hole concentration, thereby boosting the Seebeck coefficient and resistivity. The resulting power factors were low at room temperature and increased with temperature for most of the compounds (**Figure S4(a)**). However, the power factor of AgSb$_{1.021}$Te$_{2.04}$ was already high at room temperature and increased slightly with temperature, resulting in a high average value. The average power factor increased with Ag deficiencies (**Figure S4(b)**) which agrees with our and previous[29] calculations (**Figure S1(b)**). The high resistivity of the Sb-rich compounds, however, results in modest power factors.

The thermal conductivity of all samples examined was low and nearly temperature-independent (**Figure 4(e)**). The lowest thermal conductivity was found for AgSb$_{1.021}$Te$_{2.04}$ and is half that of AgSb$_{1.03}$Te$_{2.06}$. The Sb-rich compounds have a lower thermal conductivity due to their higher ρ. Furthermore, the Sb-rich compounds exhibit stereochemically active 5$s^2$ lone pairs which deform the lattice vibrations resulting in strong anharmonicity and low lattice thermal conductivity.[21] The electronic contribution ($\kappa_{el} = L_{eff}\frac{T}{\rho}$ with $L_{eff}$ as the effective Lorenz number) was calculated using the single parabolic band model in which the charge carriers are limited by acoustic deformation potential scattering[33] (**Figure S5(a)**). The electronic contribution is reciprocally dependent on the resistivity, and therefore, the Te-rich compounds exhibit a higher electronic contribution (**Figure S5(b)**) and higher total thermal conductivity (**Figure 4(f)**). The phononic contribution was calculated by subtracting the electronic part from the total thermal conductivity. $\kappa_{ph}$ ranges from 0.18 to 0.6 W m$^{-1}$ K$^{-1}$ (**Figure S5(c)**) which is similar to other reported lattice thermal conductivities.[7,18,19,20,21,22] Although slightly lower $\kappa_{ph}$ was reported for the Cd-doped compound,[7] the single parabolic band approach most likely overestimates $\kappa_{el}$ because multiple bands as well as electrons and holes contribute to the electron transport. This would explain the correlation between low electrical resistivity (**Figure 4(b)**) and low $\kappa_{ph}$ (**Figure S5(d)**). However, our combined DoE/ML approach enables us to find the optimum thermoelectric efficiency and reveal trends without requiring comprehensive insights into the complex underlying physics of the electron and phonon transport.

## 3. Conclusion

In summary, a software driven by DoE and ML was developed and successfully applied to optimize the thermoelectric performance of $AgSb_{1+x}Te_{2+y}$. The Latin Square method in the DoE framework significantly reduced the number of experiments. In just three optimization cycles and in total eight samples, the $AgSb_{1.021}Te_{2.04}$ composition with maximum ($zT_{peak}$ = 1.61±0.24) and average ($\overline{zT}$ = 1.18±0.18) figures of merit 30% higher than literature values was discovered. While other $AgSb_{1+x}Te_{2+y}$ samples contained impurity phases, limiting their thermoelectric efficiency, no evidence of a secondary phase was detected in $AgSb_{1.021}Te_{2.04}$. Furthermore, the electrical properties were enhanced by precise control of the Ag-deficiency, which agrees with the sharp peak of the theoretical power factor in the valence bands. All samples exhibited low thermal conductivity, with the lowest (~0.4 W m$^{-1}$ K$^{-1}$) for $AgSb_{1.021}Te_{2.04}$, attributed to the lack of the $Ag_2Te$ secondary phase and a moderate electronic contribution to the thermal conductivity. The software was designed to adapt the present DoE/ML approach to multi-variables problems (*e.g.*, four or more elements, multiple synthesis and process parameters). Therefore, this strategy is anticipated to create a new era in the optimization of properties, processes, and synthetic methodologies in many research areas including materials science, chemistry, biology, mechanical and chemical engineering.

## Supporting Information

Supporting Information is available upon request.

## Acknowledgments

The authors would like to thank Professor Mary Anne White (Dalhousie University) for valuable discussions. This work was supported by the Discovery Grant from the Natural Sciences and Engineering Research Council of Canada. The work was supported by the National Science Foundation Graduate Research Fellowship Program award number DGE-1848739, DGECR/284-2023, and RGPIN/4115-2023. Furthermore, JHCP would like to acknowledge the support by New Brunswick Innovation Foundation.

## Conflict of Interest

The authors declare no conflict of interest.

## Data Availability Statement

The data that support the findings of this study are available in the Supporting Information of this article. The code for the combined Machine learning and Design of Experiments approach can be found at https://github.com/JanPohls/MODEM and https://carrollpohlslab.ext.unb.ca/software/. The version of the code employed for this study is version 0.4

# References

[1] G. J. Snyder and E. S. Toberer, "Complex thermoelectric materials," *Nat. Mater*. **2008**, *7*, 105; doi: 10.1038/nmat2090.

[2] Z. M. Gibbs, F. Ricci, G. Li, H. Zhu, K. Persson, G. Ceder, G. Hautier, A. Jain, and G. J. Snyder, "Effective mass and Fermi surface complexity factor from ab initio band structure calculations," *npj Comput. Mater.* **2017**, *3*, 8; doi: 10.1038/s41524-017-0013-3.

[3] W. Chen, J.-H. Pöhls, G. Hautier, D. Broberg, S. Bajaj, U. Aydemir, Z. M. Gibbs, H. Zhu, M. Asta, G. J. Snyder, B. Meredig, M. A. White, K. Persson, and A. Jain, "Understanding thermoelectric properties from high-throughput calculations: trends, insights, and comparisons with experiment," *J. Mater. Chem. C* **2016**, *4*, 4414; doi: 10.1039/C5TC04339E.

[4] Y. Pei, H. Wang, and G.J. Snyder, "Band Engineering of Thermoelectric Materials," *Adv. Mater.* **2012**, *24*, 6125; doi: 10.1002/adma.201202919.

[5] A.F. May and G.J. Snyder, "Introduction to Modeling Thermoelectric Transport at High Temperatures Preparation, and Characterization in Thermoelectrics", CRC Press, Materials, 2012.

[6] K. Biswas, J. He, I. D. Blum, C.-I Wu, T. P. Hogan, D. N. Seidman, V. P. Dravid, and M. G. Kanatzidis, "High-performance bulk thermoelectrics with all-scale hierarchical architectures," *Nature* **2012**, *489*, 414; doi: 10.1038/nature11439.

[7] S. Roychowdhury, T. Ghosh, R. Arora, M. Samanta, L. Xie, N. K. Singh, A. Soni, J. He, U. V. Waghmare, and K. Biswas, "Enhanced atomic ordering leads to high thermoelectric performance in $AgSbTe_2$," *Science* **2021**, *371*, 722; doi: 10.1126/science.abb3517.

[8] C. Zhou, Y. K. Lee, Y. Yu, S. Byun, Z.-Z. Luo, H. Lee, B. Ge, Y.-L. Lee, X. Chen, J. Y. Lee, O. Cojocaru-Mirédin, H. Chang, J. Im, S.-P. Cho, M. Wuttig, V. P. Dravid, M. G. Kanatzidis, and I. Chung, "Polycrystalline SnSe with a thermoelectric figure of merit greater than the single crystal," *Nat. Mater.* **2021**, *20*, 1378; doi: 10.1038/s41563-021-01064-6.

[9] Q. Zhang, Q. Song, X. Wang, J. Sun, Q. Zhu, K. Dahal, X. Lin, F. Cao, J. Zhou, S. Chen, G. Chen, J. Mao, and Z. Ren, "Deep defect level engineering: a strategy of optimizing the carrier concentration for high thermoelectric performance," *Energy Environ. Sci.* **2018**, *11*, 933; doi: 10.1039/C8EE00112J.

[10] A. M. Ganose, J. Park, A. Faghaninia, R. Woods-Robinson, K. A. Persson, and A. Jain, "Efficient calculation of carrier scattering rates from first principles," *Nat. Commun*. **2021**, *12*, 2222; doi: 10.1038/s41467-021-22440-5.

[11] F. Zhou, W. Nielson, Y. Xia, and V. Ozoliņš, "Lattice Anharmonicity and Thermal Conductivity from Compressive Sensing of First-Principles Calculations," *Phys. Rev. Lett.* **2014**, *113*, 185501; doi: 10.1103/PhysRevLett.113.185501.

[12] J.-H. Pöhls, S. Chanakian, J. Park, A. M. Ganose, A. Dunn, N. Friesen, A. Bhattacharya, B. Hogan, S. Bux, A. Jain, A. Mar, and A. Zevalkink, "Experimental validation of high thermoelectric performance in $RECuZnP_2$ predicted by high-throughput DFT calculations," *Mater. Horiz.* **2021**, *8*, 209-215; doi: 10.1039/D0MH01112F.

[13] E. M. Williamson, Z. Sun, L. Mora-Tamez, and R. L. Brutchey, "Design of Experiments for Nanocrystal Syntheses: A How-To Guide for Proper Implementation," *Chem. Mater.* **2022**, *34*, 9823−9835; doi: 10.1021/acs.chemmater.2c02924.

[14] S. N. Politis, P. Colombo, G. Colombo, and D. M. Rekkas, "Design of experiments (DoE) in pharmaceutical development," *Drug Dev. Ind. Pharm.* **2017**, *43*, 889–901; doi: 10.1080/03639045.2017.1291672.

[15] B. Cao, L. A. Adutwum, A. O. Oliynyk, E. J. Luber, B. C. Olsen, A. Mar, and J. M. Buriak, "How To Optimize Materials and Devices via Design of Experiments and Machine Learning: Demonstration Using Organic Photovoltaics," *ACS Nano* **2018**, *12*, 7434−7444; doi: 10.1021/acsnano.8b04726.

[16] J. Gilman, L. Walls, L. Bandiera, and F. Menolascina, "Statistical Design of Experiments for Synthetic Biology," *ACS Synth. Biol.* **2021**, *10*, 1 – 18; doi: 10.1021/acssynbio.0c00385.

[17] V. Jovovic and J. P. Heremans, "Measurements of the energy band gap and valence band structure of $AgSbTe_2$," *Phys. Rev. B*, **2008**, *77*, 245204; doi: 10.1103/PhysRevB.77.245204.

[18] R. Mohanraman, R. Sankar, K. M. Boopathi, F.-C. Chou, C.-W. Chu, C.-H. Lee, and Y.-Y. Chen, "Influence of In doping on the thermoelectric properties of an $AgSbTe_2$ compound with enhanced figure of merit," *J. Mater. Chem. A* **2014**, *2*, 2839; doi: 10.1039/c3ta14547f.

[19] B. L. Du, H. Li, and X. F. Tang, "Enhanced thermoelectric performance in Na-doped p-type nonstoichiometric $AgSbTe_2$ compound," *J. Alloys Compd.* **2011**, *509*, 2039–2043; doi: 10.1016/j.jallcom.2010.10.131.

[20] B. Du, H. Li, J. Xu, X. Tang, and C. Uher, "Enhanced Figure-of-Merit in Se-Doped P-Type $AgSbTe_2$ Thermoelectric Compound," *Chem. Mater.* **2010**, *22*, 5521–5527; doi: 10.1021/cm101503y.

[21] S. Roychowdhury, R. Panigrahi, S. Perumal, and K. Biswas, "Ultrahigh Thermoelectric Figure of Merit and Enhanced Mechanical Stability of p-type AgSb$_{1-x}$Zn$_x$Te$_2$," *ACS Energy Lett.* **2017**, *2*, 349−356; doi: 10.1021/acsenergylett.6b00639.
[22] J. K. Lee, M.-W. Oh, B. Ryu, J. E. Lee, B.-S. Kim, B.-K. Min, S.-J. Joo, H.-W. Lee, and S.-D. Park, "Enhanced thermoelectric properties of AgSbTe$_2$ obtained by controlling heterophases with Ce doping," *Sci. Rep.* **2017**, *7*, 4496; doi: 10.1038/s41598-017-04885-1.
[23] M. D. Nielsen, V. Ozolins, and J.-P. Heremans, "Lone pair electrons minimize lattice thermal conductivity," *Energy Environ. Sci.* **2013**, *6*, 570−578; doi: 10.1039/C2EE23391F.
[24] J. Ma, O. Delaire, A. F. May, C. E. Carlton, M. A. McGuire, L. H. VanBebber, D. L. Abernathy, G. Ehlers, T. Hong, A. Huq, W. Tian, V. M. Keppens, Y. Shao-Horn, and B. C. Sales, "Glass-like phonon scattering from a spontaneous nanostructure in AgSbTe$_2$," *Nat. Nanotechnol.* **2013**, *8*, 445−451; doi: 10.1038/nnano.2013.95.
[25] D. T. Morelli, V. Jovovic, and J.-P. Heremans, "Intrinsically minimal thermal conductivity in cubic I-V-VI$_2$ semiconductors," *Phys. Rev. Lett.* **2008**, *101*, 035901; doi: 10.1103/PhysRevLett.101.035901.
[26] L.-H. Ye, K. Hoang, A. J. Freeman, S. D. Mahanti, J. He, T. M. Tritt, and M. G. Kanatzidis, "First-principles study of the electronic, optical, and lattice vibrational properties of AgSbTe$_2$," *Phys. Rev. B* **2008**, *77*, 245203; doi: 10.1103/PhysRevB.77.245203.
[27] J. Ma, O. Delaire, E. D. Specht, A. F. May, O. Gourdon, J. D. Budai, M. A. McGuire, T. Hong, D. L. Abernathy, G. Ehlers, and E. Karapetrova, "Phonon scattering rates and atomic ordering in Ag$_{1-x}$Sb$_{1+x}$Te$_{2+x}$ ($x$=0,0.1,0.2) investigated with inelastic neutron scattering and synchrotron diffraction," *Phys. Rev. B* **2014**, *90*, 134303; doi: 10.1103/PhysRevB.90.134303.
[28] J. Cao, J. Dong, K. Saglik, D. Zhang, S. F. D. Solco, I. J. W. J. You, H. Liu, Q. Zhu, J. Xu, J. Wu, F. Wei, Q. Yan, and A. Suwardi, "Non-equilibrium strategy for enhancing thermoelectric properties and improving stability of AgSbTe$_2$," *Nano Energy* **2023**, *107*, 108118; doi: 10.1016/j.nanoen.2022.108118.
[29] W. Szczypka and A. Koleżyński, "Self-compensating defects in AgSbTe$_2$ from first principles studies," *J. Alloys Compd.* **2019**, *787*, 1136e1142; doi: 10.1016/j.jallcom.2019.02.082.
[30] https://github.com/JanPohls/MODEM; Date: August 30th, 2024.
[31] H.-j. Wu, S.-w. Chen, T. Ikeda, and G. J. Snyder, "Reduced thermal conductivity in Pb-alloyed AgSbTe$_2$ thermoelectric materials," *Acta Mater.* **2012**, *60*, 6144–6151; doi: 10.1016/j.actamat.2012.07.057.
[32] Y. Chen, M. D. Nielsen, Y.-B. Gao, T.-J. Zhu, X. Zhao, and J. P. Heremans, "SnTe–AgSbTe$_2$ Thermoelectric Alloys," *Adv. Energy Mater.* **2012**, *2*, 58–62; doi: 10.1002/aenm.201100460.
[33] J.-H. Pöhls and Y. Mozharivskyj, "TOSSPB: Thermoelectric optimization based on scattering-dependent single-parabolic band model," *Comput. Mater. Sci*. **2022**, *206*, 111152; doi: 10.1016/j.commatsci.2021.111152.